\documentclass[twocolumn,showpacs,preprintnumbers,amsmath,amssymb]{revtex4}

\usepackage{graphicx}
\usepackage{dcolumn}
\usepackage{bm}

\begin{document}


\title{Mass-renormalized electronic excitations at ($\pi$, 0) \\ in
the superconducting state of $Bi_{2}Sr_{2}CaCu_{2}O_{8+\delta}$}

\author{A. D. Gromko$^{1}$, A. V. Fedorov$^{1,2}$, Y. -D.
Chuang$^{1,2}$, J. D. Koralek$^{1}$,
Y. Aiura$^{3}$, Y. Yamaguchi$^{3}$, K. Oka$^{3}$, Yoichi
Ando$^{4}$, D. S. Dessau$^{1}$}

\affiliation{$^{1}$Department of Physics, University of Colorado,
Boulder, Colorado 80309-0390}

\affiliation{$^{2}$Advanced Light Source, Lawrence Berkeley
National Laboratory, Berkeley, CA 94720}

\affiliation{$^{3}$National Institute for Advanced Industrial
Science and Technology (AIST), AIST Tsukuba Central 2, 1-1-1
Umezono, Tsukuba, Ibaraki 305-8568, Japan}

\affiliation{$^{4}$Central Research Institute of Electric Power
Industry (CRIEPI), 2-11-1 Iwato-Kita, Komae, Tokyo 201-8511,
Japan}

\date{\today}

\begin{abstract}

Using high-resolution angle-resolved photoemission spectroscopy on
$Bi_{2}Sr_{2}CaCu_{2}O_{8+\delta}$,
we have made the first observation of a mass renormalization or
"kink" in the E vs. $\vec k$ dispersion relation
localized near $(\pi, 0)$.  Compared to the kink observed
along the nodal direction, this new effect is clearly stronger, appears
at a lower energy near 40 meV, and is only present
in the superconducting state.  The kink energy scale defines a
cutoff below which well-defined quasiparticle excitations occur.
This effect is likely due to coupling to a bosonic excitation, with
the most plausible candidate being the magnetic resonance mode observed
in inelastic neutron scattering.

\end{abstract}

\maketitle



One of the most pressing scientific questions in condensed matter
physics concerns the mechanism that binds two electrons into a
Cooper pair in high temperature superconductors (HTSCs). It is
known that these pairs have d-wave symmetry, with the
superconducting gap, $\Delta(\vec{k})$, having a maximum value at
the $(\pi, 0)$ point of the two-dimensional Brillouin zone, and
zero gap $45^{\circ}$ away along the $(\pi, \pi)$
line~\cite{Shen:1993,Tsuei:2000}. In conventional superconductors
it is known that the s-wave pairing is mediated by the
electron-phonon interaction. Most forms of this interaction are
isotropic in $\vec k$-space, and are generally unfavorable for the
d-wave pairing state found in HTSCs. More likely important for
HTSC is a boson that couples strongly to the $(\pi, 0)$ electrons
at the d-wave maximum. Here we show direct observations that
electrons at $(\pi, 0)$ are strongly coupled to a bosonic
excitation, with indications that this boson consists of the
famous magnetic resonance mode long observed in inelastic neutron
scattering (INS) experiments~\cite{Bourges:1998}. We find that
this interaction only exists in the superconducting state and is a
large effect, strongly renormalizing the E vs. $\vec k$ dispersion
of the low-energy electronic states.

In the many-body language of solid state physics, the electron
self-energy, $\Sigma(\vec k,w)$, contains the information of the
interactions or correlation effects which "dress" the free
electrons in a solid to make quasiparticles. This dressing
renormalizes the dispersion of electrons near the Fermi energy,
giving them an enhanced mass or flatter E vs. $\vec k$ dispersion.
At high energies (greater than the energy of the boson being
coupled to), the dispersion returns to its bare value, giving the
dispersion a "kink". The energy scale and strength of the kink are
thus related to the boson energy and coupling strength
respectively. Much effort has recently been put forth to study
dispersion kinks in relation to HTSC, with the hopes of uncovering
the phonons or other bosonic modes which couple to the
near-$E_{F}$ electrons. ARPES is the only direct method for this,
with essentially all efforts to date focused on the nodal region
of the Brillouin
zone~\cite{Bogdanov:2000,Kaminski:2001,Lanzara:2001,Johnson:2001},
where the d-wave superconducting gap goes to zero. Measurements
near the $(\pi, 0)$ point of the Brillouin zone, where much of the
exotic physics of HTSC such as the maximum in the d-wave
superconducting gap~\cite{Shen:1993,Tsuei:2000} and normal state
pseudogap~\cite{Ding:1996,Loeser:1996} occur, have been limited.
The main reason for the lack of kink measurements near $(\pi, 0)$
to date are the large widths of spectral features there. While
these widths affect the accuracy of determining peak positions,
more importantly they can hinder a deconvolution of the
bilayer-split bands into their bonding and antibonding components,
disallowing measurements of their dispersion.  By overdoping high
quality single-crystalline $Bi_{2}Sr_{2}CaCu_{2}O_{8+\delta}$
samples, we have obtained very sharp spectral features near $(\pi,
0)$ and for the first time have been able to accurately deconvolve
the bilayer splitting as well as superstructure
effects~\cite{Chuang:2001,Chuang:preprint} with similar work done
by Feng et al.~\cite{Feng:2001}. The ability to clearly resolve
these separate features is one of the key steps that enabled the
new experiments reported here.
%
%
%
\begin{figure}
\includegraphics[width=3.5in]{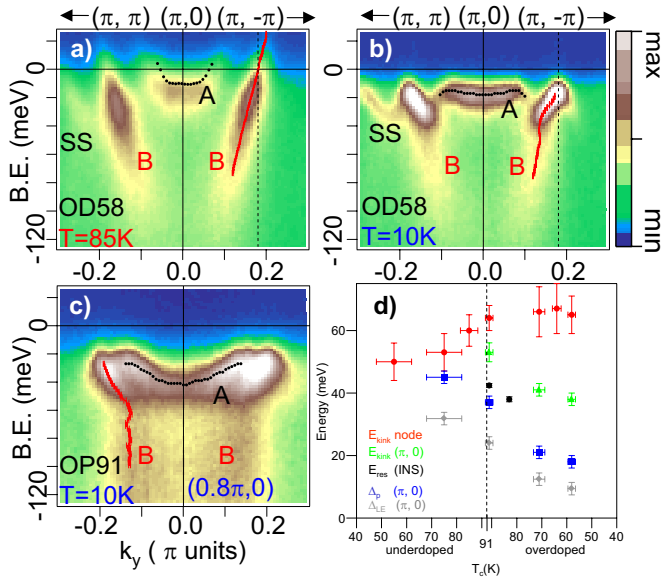}
\caption{\label{fig1} (a) Normal and (b) superconducting state
ARPES data from sample OD58, showing dispersion of the
bilayer-split bonding (B) and antibonding (A) bands along the
$(\pi,-\pi)-(\pi,0)-(\pi,\pi)$ symmetry line (see figure 3(e)).
(c) Superconducting state data taken on sample OP91, along
$(0.8\pi,-\pi)-(0.8\pi,0)-(0.8\pi,\pi)$.  (d) A summary of the
energy scales of the ARPES kinks and the INS resonance, for both
underdoped (left) and overdoped (right) samples. $E_{kink}(node)$
is from our own unpublished data~\cite{Gromko:preprint} and
$E_{res}(INS)$ is from ref.~\cite{He:1999}.}
\end{figure}
%
%
%

We performed high-resolution ARPES experiments on
$Bi_{2}Sr_{2}CaCu_{2}O_{8+\delta}$ (Bi2212) samples over a wide
range of oxygen concentrations. Samples are labelled with a
convention based on the transition temperature ($T_c$), i.e. an
overdoped (OD) Bi2212 sample with $T_c=58K$ is referred to as
OD58. The same convention is used for optimal (OP) and underdoped
(UD) samples. All ARPES measurements were taken at beamline 10.0.1
of the Advanced Light Source, Berkeley, and at beamline 5-4 of the
Stanford Synchrotron Radiation Laboratory using SES 200 electron
spectrometers. The experiments were done using 20eV photons, with
a combined experimental energy resolution of 12meV, and a momentum
resolution better than $0.01\pi/a$ (where a is the Bi2212 lattice
constant) along the entrance slit to the spectrometer.

Figures 1(a)-(c) show raw data near the $(\pi, 0)$ region (see
figure 3(e)) taken on samples OD58 and OP91. A salient feature of
the data is the clear resolution of two bands, the higher binding
energy bonding band (B) and the lower binding energy antibonding
band (A), plus some weak superstructure bands (SS) due to the
extra periodicity induced by the Bi-O plane lattice mismatch.
Superimposed on top of the B band we show the peak positions (red)
determined from Lorentzian fitting of Momentum Distribution Curves
(MDCs), which are cuts in momentum space at constant energy.  For
this data the MDCs display a simple Lorentzian lineshape, allowing
us to accurately determine the dispersion. Error bars from the
fits are included, but are so small that they are essentially
invisible. We also include the dispersion of the A band (black
dots), extracted from the sharp low-energy peak in the Energy
Distribution Curves (EDCs). The normal state dispersion shown in
panel (a) is seen to be nearly linear and featureless in the
energy range displayed. Upon cooling the sample to 10K (panel
(b)), the dispersion as well as near-$E_{F}$ spectral weight are
radically changed.  First, the features do not reach $E_{F}$
because of the opening of the superconducting gap $\Delta$. In
addition to the gap opening, there is a clear kink in the
dispersion around 40 meV. Although for sample OP91 the spectral
features are broader due to decreased doping (panel (c)), the data
show a similar effect.

The temperature dependence of the $(\pi, 0)$ kink can be seen in
figure 2(a), where the MDC-derived bonding band dispersion from
sample OD71 is shown at a series of temperatures taken on cooling
through $T_c$.  The black dotted line is a linear fit to the
highest temperature ($T=85K$) data. Since this dispersion is
featureless, it is considered to be the non-interacting
dispersion. A kink near 40 meV opens up as the temperature is
lowered, in addition to low energy changes that are associated
with the opening of the superconducting gap~\cite{SCMDCcomment}.
To highlight the temperature dependent effects we subtract the
superconducting state MDC dispersion from the normal state
dispersion, shown in the inset to panel (a). In this way we
extract $Re\Sigma$ from the data. Panel (b) shows the maximum
point of each $Re\Sigma$ curve from the inset in panel (a) as a
function of temperature (red), which is seen to have an onset at
or very near $T_{c}$ = 71K. In addition, we also plot the
superconducting gap $\Delta_{LE}$ as a function of temperature
(blue) extracted from the identical data set by simply measuring
the shift of the midpoint of the EDC leading edge from $E_{F}$. We
plot $\Delta_{LE}(T)$ in lieu of the gap $\Delta_{p}(T)$,
determined from EDC peak positions, because a reliable
determination of $\Delta_{p}(T)$ is made difficult by thermal
broadening. We see that the maximum in $Re\Sigma$, which provides
a measure of the coupling strength, tracks the opening of the
superconducting gap, making it clear that the two are closely
linked.
%
%
%
\begin{figure} 
\includegraphics[width=3.5in]{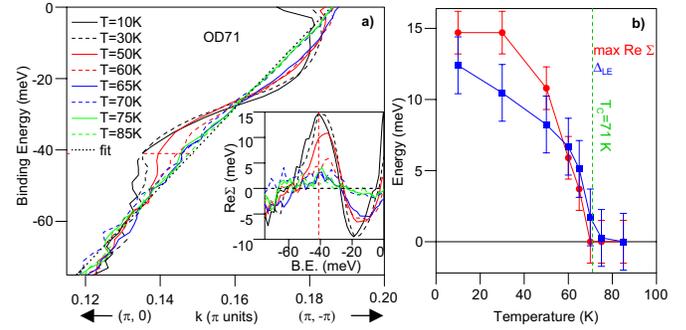}
\caption{\label{fig2} (a) The MDC bonding-band dispersion from
sample OD71 along $(\pi, 0)- (\pi, -\pi)$. The black dotted line is a
linear fit to
the $T=85K$ curve. The inset to panel (a) shows the temperature
dependence of $Re\Sigma$ determined using the linear fit. The red
dashed line indicates the kink energy of $\approx 40meV$. (b) The
temperature dependence of the maximum in $Re\Sigma$ (red circles)
and the superconducting gap $\Delta_{LE}(T)$ (blue squares).}
\end{figure}
%
%
%

One may imagine that the $(\pi, 0)$ kink is a byproduct of the
opening of the superconducting gap.  In particular, if we consider
damping of the system due to electron-hole excitations, then each
branch will see a turn on at the gap energy $\Delta$, leading to a
step in the damping rate $Im\Sigma$ near an energy $3\Delta$. This
will in turn introduce structure into $Re\Sigma$ (as seen by a
Kramers-Kr\"onig analysis), and could be imagined as the origin of
the kink.  This is somewhat attractive because the temperature as
well as $\vec k$-space dependence to first order appear reasonable
- it should turn on at $T_{c}$ and the d-wave nature of $\Delta$
may cause some localization of the effect near $(\pi, 0)$.
However, as the temperature is raised towards $T_{c}$, the energy
of the kink is observed to decrease slowly, staying at a sizeable
finite value (figure 2(a)).  In contrast, the $3\Delta$ model
would predict that the kink energy should decrease to zero at
$T_{c}$, just as $\Delta$ does~\cite{pseudogapcomment}.

Figures 3(a)-(d) show the $\vec k$-dependence of the ARPES kink,
measured on sample OD71.  Data were taken at four momentum slices
parallel to the standard $(\pi, 0)$ cuts of figures 1(a)-(b),
centered around the $\vec k$ values of $(\pi, 0)$, $(0.9\pi, 0)$,
$(0.8\pi, 0)$, and $(0.7\pi, 0)$ (blue bars, panel (e)). Panels
(a)-(d) show the MDC dispersion of the bonding band from portions
of these cuts in both the normal (red) and superconducting (blue)
states. From the progression we see that the kink weakens
dramatically as we move away from the $(\pi, 0)$ point, such that
it is barely visible in the $0.7\pi$ cut of panel (d).

There is another key physical property discussed in the HTSC
literature that has a similar temperature and momentum dependence
as the kink, namely the magnetic resonance mode observed in
inelastic neutron scattering (INS) experiments. The resonance mode
only occurs below $T_{c}$~\cite{Fong:1999}, or in underdoped
samples below $T^{*}$~\cite{Dai:1999}. It has a characteristic
energy near 40meV for optimally doped samples and is centered at a
wavevector $\vec Q=(\pi, \pi)$.
The open black circles of figure 4(f) show the resonance mode
intensity as a function of momentum transfer $\vec Q$ for an OD83
Bi2212 sample, extracted from the INS studies of He et
al.~\cite{He:1999}. The mode half-intensity points are at
approximately $\vec Q=0.66(\pi, \pi)$ and $1.33(\pi, \pi)$ along
this cut (green dashed lines, panel (f)), and are expected to be
roughly isotropic in $\vec Q$-space~\cite{Mook:1998}. We show
these half-intensity points schematically on the Brillouin zone in
panel (e) as green circles centered around $(\pi, 0)$ and $(0,
\pi)$, according to the standard belief that the $(\pi, \pi)$ mode
connects these points since they have the largest near-$E_{F}$
electron density and are separated by a vector of $(\pi, \pi)$
~\cite{Ioffe:1998}. The circles are drawn so that the closest
(furthest) edges are connected by a $\vec Q$ of $0.66(\pi, \pi)$
$(1.33(\pi, \pi))$.  In this way, if we were to plot circles
representative of the endpoints of the bottom axis in panel (f),
these circles would just touch each other.
%
%
%
\begin{figure}
\includegraphics[width=3.5in]{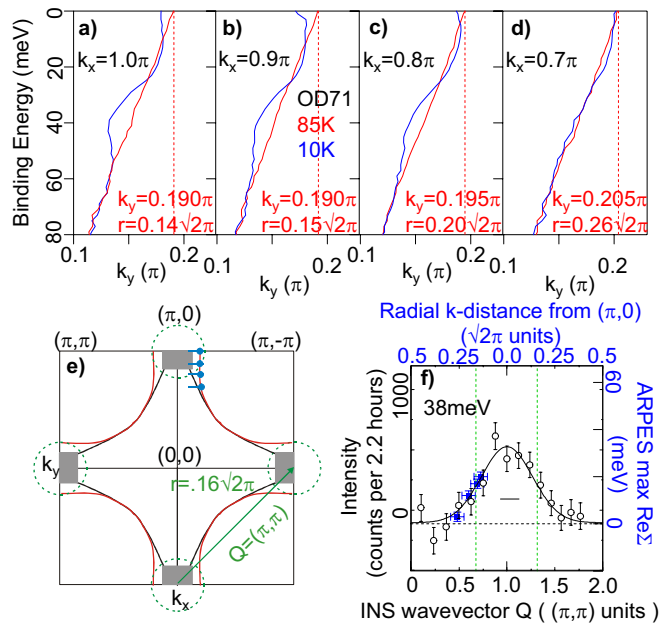}
\caption{\label{fig3} (a)-(d) The MDC-derived bonding-band
dispersion extracted from momentum cuts on sample OD71 (shown as
blue bars in panel (e)), at 85K (red) and 10K (blue). The red
dashed lines mark the FS crossing locations, which are
also labelled on each panel along with the radial distance from
$(\pi, 0)$. (e) The Fermi surface with
INS wavevectors (green) connecting the $(\pi, 0)$ and $(0, \pi)$
points. Momentum transfer from the surface of one green circle to
another represents the INS half-intensity values. (f) INS
intensity vs. $\vec Q$ from He et al. (33). Overlaid with this is
the maximum in $Re \Sigma$ (blue squares) versus the radial
distance from $(\pi, 0)$ (top axis). }
\end{figure}
%
%
%

We converted the $\vec k$-space values of
the Fermi surface crossing locations (panel (e), blue dots) into a
radial $\vec k$-distance from the $(\pi, 0)$ points in units of
$\surd 2 \pi$ , listed in panels (a) - (d). The top axis of panel
(f) shows the maximum in $Re \Sigma$ plotted versus radial
distance from $(\pi, 0)$, with the edges of the plot again
corresponding to the situation where circles centered at $(\pi, 0)$
just touch. Since we do not have kink
data exactly at the $(\pi, 0)$ point, we scale the vertical axes
to match for the $(\pi, 0.19\pi)$ point. The excellent agreement
between the neutron and ARPES intensities makes an intimate
relationship between the kink and the magnetic mode highly
plausible.

The trends in the ARPES spectral weight also support the notion of
coupling to a boson such as the magnetic mode. In particular, the
superconducting state ARPES peaks are seen to be sharp and intense
for binding energies below the kink energy (figure 1(b)), whereas
they are strongly damped for binding energies above the kink
energy, as well as for all energies in the normal state (figure
1(a)). This is expected within the bosonic coupling picture
because at low energies, only virtual excitations of the boson are
present, which renormalize the electronic excitations.  At high
energies, real excitations of the boson will be allowed, causing a
rapid turn-on of the damping effect. The peak broadening in the
normal state, which is believed to represent its
"non-quasiparticle-like" nature, has also received intense
attention~\cite{Laughlin:1999,Shen:1997,Sawatzky:1989}.
Interestingly, at high binding energies (100 meV and above), the
spectral weight is essentially identical in the normal and
superconducting states.  A possible explanation for this is that
above $T_{c}$ the magnetic excitations are greatly smeared out in
energy but retain a similar $\vec Q$-dependence, as neutron data
is beginning to show~\cite{Bourges:2000}.

 From the above data, the differences between the $(\pi, 0)$ kink
and the well-studied nodal kink can be summarized as follows: 1)
The $(\pi, 0)$ kink strength is strongly temperature dependent
(figure 2), while the nodal kink is roughly temperature
independent~\cite{Lanzara:2001,Johnson:2001}. 2) The $(\pi, 0)$
kink is at a significantly lower energy scale than the nodal kink
(figure 1(d)) and has a different momentum dependence. 3) The
strength of the $(\pi, 0)$ kink (or coupling strength) is much
stronger than the nodal kink~\cite{Gromko:preprint}. These points
make a strong case that the two kinks are separate entities. The
very strong temperature dependence of the $(\pi, 0)$ kink, as well
as the tight localization of the kink in k-space make it unlikely
that electron-phonon coupling is responsible for the $(\pi, 0)$
kink, although it does remain a possibility for the nodal
kink~\cite{Lanzara:2001}. Another possibility for the origin of
the nodal kink is coupling to the local magnetic susceptibility,
which is observed in $\vec Q$-integrated INS measurements and
which occurs with a slightly higher energy than the magnetic
resonance mode~\cite{Bourges:1998}.

Figure 1(d) shows a comparison of the energy scales of the ARPES kink
at $(\pi, 0)$ (green triangles), the INS
resonance~\cite{He:1999}(black circles), and the nodal kink from
our own data~\cite{Gromko:preprint}(red circles). Following the
analogy of $\alpha^{2}F(\omega)$ oscillations in strongly
electron-phonon coupled conventional s-wave superconductors, we
may expect the $(\pi, 0)$ kink to be observed at an energy equal
to $\Delta+\omega_{R}$, where $\Delta$ is the superconducting gap
and $\omega_{R}$ is the neutron resonance mode energy. In the
presence of a varying d-wave gap with a broadened edge, there is
no a-priori portion of the gap or gap-edge from which to reference
the mode energy.  If we choose the superconducting state EDC peak
position $\Delta_{p}$ at $(\pi, 0)$ (blue squares), the energies
$\Delta_{p}+\omega_{R}$ will clearly be larger than the kink
energy scale, though if we choose the leading edge half-maximum
point $\Delta_{LE}$ at $(\pi, 0)$ (gray diamonds), the agreement
between $\Delta_{LE}+\omega_{R}$ and the kink energy scale is much
better. Regardless of whether this should be considered an
agreement of energy scales, it is not surprising that the
situation should be more complicated in the cuprates than in
conventional superconductors.  For one, the resonance mode to
which the electrons appear to be coupling is not yet understood,
even at the basic level of whether it corresponds to
particle-particle~\cite{Demmler:1995} or
particle-hole~\cite{Eschrig:2000} excitations.

%
%
%
%
%
%

Since the magnetic mode is electronic in origin and only occurs in
the paired state, there is a likelihood of a strong feedback
effect that may change the observed energy scale. This contrasts
with the traditional electron-phonon coupling system, where
lattice dynamics are minimally affected by the onset of pairing
correlations. In some limits, the energy scale of electronic
coupling to the magnetic mode may be similar to the $3\Delta$ case
discussed earlier~\cite{Abanov:2000}, as the magnetic mode is
often considered to be an electron-hole excitation, albeit one
with highly defined constraints in momentum. Finally, we note that
the strong coupling near the $(\pi, 0)$ point makes vertex
corrections much more important~\cite{Zhang:pcomm}, and these as
well as the possibility of excitonic effects~\cite{Abanov:2000}
may lower the energy of the kink.

In conclusion, we have made the first observation of a kink in the
ARPES spectra of HTSC cuprates near the critical $(\pi, 0)$ point
of the zone, where the pairing is strongest.  The temperature and
$\vec k$-dependence of this kink suggest it originates from strong
coupling of electrons to the collective magnetic resonance mode
long observed in inelastic neutron scattering experiments.  This
increases the possibility that the magnetic
mode acts as the "glue" that couples the electrons together within
a Cooper pair.

We acknowledge sample preparation help from M. Varney, beamline
support from X.J. Zhou, P. Bogdanov, Z. Hussain and D.H. Liu, and
helpful discussions with G. Aeppli, A. Chubukov, C. Kendziora, A.
Millis, P. Lee, D. Pines, D. Scalapino, J. Schmallian, Z-X. Shen,
and S.C. Zhang.  We gratefully acknowledge the help of R. Goldfarb
at NIST for the use of the SQUID magnetometer. This work was
supported by the NSF Career-DMR-9985492 and the DOE
DE-FG03-00ER45809.  ALS and SSRL are operated by the DOE, Office
of Basic Energy Sciences.


\begin{references}
\bibitem{SCMDCcomment} An artifact of the MDC analysis is that inside
the gap region
($\approx 0-25 meV$), where there is no real quasiparticle peak,
the MDC method finds a peak due to the finite energy width of the
spectra. We thus focus on the energy range between 25 and 75 meV,
where the kink near 40 meV shows a continuous evolution as a
function of temperature.
\bibitem{pseudogapcomment} In underdoped samples the gap may remain
finite up to and across $T_{c}$, a phenomenon not found in
overdoped samples.
\end{references}

\end{document}